# Superconductivity in the high-pressure tetragonal phase of $UTe_2$

Yuhang Deng[1], Gabriel Mee[1], Tyler Wannamaker[1], Keke Feng[1],

Mingyu Xu[2], Weiwei Xie[2], M. Brian Maple[1,*]

[1]Department of Physics, University of California San Diego, La Jolla, California 92903, USA

[2]Department of Chemistry, Michigan State University, East Lansing, Michigan 48824, USA

[*]Corresponding author: M. Brian Maple, mbmaple@ucsd.edu

## Abstract

Electrical transport and magnetic measurements have been made on $UTe_2$ under pressure $P$ up to ~16 GPa to determine the superconducting transition temperature $T_c$ vs $P$ phase diagram in the high-pressure tetragonal phase. Superconductivity emerges near ~5 GPa, coincident with the orthorhombic – tetragonal phase transition; in the tetragonal phase, $T_c$ reaches a maximum value of ~4 K at 6 GPa and then decreases with $P$ and appears to vanish near 18 GPa. Tetragonal $UTe_2$ has a relatively small upper critical field $H_{c2}(0) \approx 1.2$ T at 5.3 GPa, smaller than the Pauli paramagnetic limit, and is orbitally limited with a coherence length $\xi_{tetra} \approx 16.5$ nm. This small value of $H_{c2}(0)$ favors more conventional superconductivity; in contrast, the large values of $H_{c2}(T)$ for orthorhombic $UTe_2$ exceed the Pauli paramagnetic limit in all three crystallographic directions and have been attributed to unconventional superconductivity, widely believed to involve spin-triplet pairing. The temperature–pressure phase diagram of $UTe_2$ shows a striking dichotomy: a narrow, fragile, unconventional superconducting region in the orthorhombic phase vs a broad, robust, and more conventional superconducting dome in the tetragonal phase. This dichotomy is consistent with the proposal that U-dimers, present (absent) in the orthorhombic (tetragonal) phase, may play a role in spin-triplet superconductivity of orthorhombic $UTe_2$. In the tetragonal phase, the normal-state electrical resistivity $\rho(T)$ exhibits metallic behavior with a "knee" at 200 – 240 K that depends weakly on $P$ and most likely marks the onset of a transition to a magnetically ordered phase that coexists with superconductivity.

## Significance Statement

Understanding the origin and nature of conventional and unconventional superconductivity remains a central challenge in strongly correlated electron physics. In $UTe_2$, we mapped out the temperature vs pressure phase diagram of a new superconducting state hosted in the high-pressure tetragonal phase that exhibits a broad superconducting dome, small upper critical fields, and weaker electronic correlations compared with the unconventional superconductivity in orthorhombic $UTe_2$. The striking contrast between the two phases suggests that structural changes and uranium bonding configurations strongly influence the superconducting pairing mechanism and the underlying electronic correlations. Our results establish $UTe_2$ as a rare platform for exploring the interrelation between distinct superconducting regimes tuned by pressure within a single material.

## Introduction

The recently discovered heavy-fermion compound uranium ditelluride ($UTe_2$) with $T_c$ = 2.1 K has gathered intense interest due to its extraordinary superconducting properties at ambient pressure [Ran19, Aoki19], magnetic field reentrant and field polarized superconductivity ($SC_{FP}$) [Knebel19, Ran19b, Frank24, Moir25], and complex interplay between ferromagnetic spin fluctuations [Sundar19, Ambika22], antiferromagnetic (AFM) spin fluctuations [Duan20, Knafo21, Duan21, Raymond21, Butch22] and unconventional superconductivity. Its potential realization of spin-triplet, chiral, odd-parity superconductivity positions $UTe_2$ as a strong candidate for topological superconductivity, offering promising avenues for quantum technologies [Ran19, Lin20, Li25, Ishizuka19, Shishidou21, Hayes21]. However, despite extensive studies, a comprehensive understanding of its superconducting and magnetic properties, electronic states, and crystal structures remains elusive, especially under high pressure where limited experimental probes could be applied.

Under applied pressure ($P$) below 4 GPa and in high magnetic fields, $UTe_2$ undergoes a series of electronic and magnetic transitions. At ambient pressure and in high magnetic field, $UTe_2$ has three superconducting phases: $SC_1$ below ~20 T, $SC_{HF}$ between 20 and 35 T near the *b*-axis, and $SC_{FP}$ above 35 T that exists in a narrow angular region in the *bc* plane inside the field polarized region [Wu24]. When $P$ > 0.3 GPa, another superconducting phase, the high-pressure-zero-field phase $SC_2$, appears and coexists with $SC_1$ with a higher $T_c$ [Braithwaite19, Thomas20]. It was later concluded that the $SC_2$ and $SC_{HF}$ phases have the same identity by means of specific heat measurements as a function of $H$, $P$ and $T$ which show that the two phases evolve continuously into one another [Vasina25]. Under higher pressure ($P_c$ > ~1.7 GPa), $SC_1$ and $SC_2$ are suppressed and replaced by long-range AFM order, revealed by neutron diffraction [Knafo25], ac calorimetry, and electrical transport measurements [Thomas20]. The emergence of this AFM order is likely due to the pressure-enhanced AFM fluctuations near $P_c$ [Knafo25] and seems to occur simultaneously with a U valence change towards $U^{4+}$ [Thomas20, Wilhelm23].

When $P$ is higher than ~4 GPa, $UTe_2$ undergoes a transition from an orthorhombic (space group *Immm*) to a higher-symmetry tetragonal crystal structure (*I*4/*mmm*), accompanied by a sudden volume collapse with an increase in the nearest U-U separation [Huston22, Honda23] and a change in the uranium electronic configuration towards a larger fraction of the $5f^{\,3}$ configuration [Wilhelm23, Deng24] revealed by x-ray spectroscopies. In the tetragonal phase, the electrical transport behavior of $UTe_2$ is characterized by a knee at $T^{**}$ ≈ 230 K in electrical resistivity ρ vs temperature $T$ curves [Honda23], which was suggested to represent as a crossover from localized 5*f*-electron behavior to a coherent array of 5*f*-electrons [Honda23], or a transition to a magnetically ordered state [Thebault24]. Moreover, bordering the AFM phase at lower pressures in the orthorhombic phase, a new superconducting region emerges in the tetragonal phase with weaker electronic correlations and smaller upper critical fields $H_{c2}$, revealing a new type of superconductivity that is distinctly different from the low-pressure heavy-fermion unconventional superconductivity [Honda23]. Interestingly, it was claimed that possible metamagnetic transitions appear in tetragonal $UTe_2$ in high magnetic fields applied nearly along the *c*'-axis of the tetragonal phase, or tilted at 30° from the *b*-axis towards the *c*-axis of the orthorhombic phase [Thebault24], exactly where the orthorhombic $SC_{FP}$ phase is accommodated after the metamagnetic transition induced by high fields. Accessing pressures up to 16 GPa, we combined electrical resistivity, AC magnetic susceptibility, and structural insights to track the evolution of the properties of $UTe_2$ across the orthorhombic–tetragonal phase transition and uncover the more complete superconducting phase diagram of the tetragonal phase of $UTe_2$.

## Experimental methods

Single crystals of $UTe_2$ were synthesized by chemical vapor transport (CVT) [Ran19] or molten salt flux liquid transport (MSFLT) [Aoki24]. No impurity phases were detected according to powder X-ray diffraction (XRD) using a Rigaku MiniFlex600 diffractometer. Superconductivity at ambient conditions was confirmed by specific heat and electrical resistivity measurements. The MSFLT-grown $UTe_2$ single crystals have a higher $T_c$ (~2.1 K) compared to the CVT-grown crystals with a $T_c$ of ~2.0 K, consistent with previous report by Aoki [Aoki24]. High-pressure XRD measurements were conducted to study the crystal structure of $UTe_2$ upon increasing and decreasing pressure, with in-depth technical descriptions in the Supporting Information [SI].

Diamond anvil cells (DACs) from Almax easyLab and DAC Tools were used to measure the electrical resistance ($R$) and AC magnetic susceptibility ($\chi$), respectively, of $UTe_2$ under high pressure. Several measurements were made, among which Run 3 and Run 4 yielded good data. In runs 3 and 4, the resistance of the $UTe_2$ sample was measured under pressure using steatite as the pressure transmitting medium (PTM) by means of the van der Pauw 4-point method. Part way through Run 3, one of the electrical leads lost contact with the sample, so a 3-point configuration was used for measurements at some pressures. For AC $\chi(T)$ runs, hardened CuBe gaskets were used to minimize the magnetic background, a mixture of Fluorinert 77:70 = 1:1 was employed as the PTM, and the flux expelled by the superconductor was detected using a mutual inductance mini coil with a primary and two oppositely wound secondary coils, one of which contained the sample. The fluorescence of ruby was measured to determine the room-temperature pressure of the $UTe_2$ sample. Additionally, a flake of lead (Pb) was placed along with the sample in the AC $\chi(T)$ experiments to estimate the pressure from the known $T_c$ of Pb [Eiling81]. For the $R$ measurements, pressures at low temperature were calibrated with $T_c$ of Pb measured in separate experiments. A customized $^4$He cryostat was used to reach temperatures as low as ~1.2 K by pumping on liquid helium with a Stokes pump. A Physical Property Measurement System (PPMS) DynaCool provided magnetic fields for magnetoresistance measurements.

## Results and Discussion

The two AC $\chi(T,P)$ runs revealed a diamagnetic signal at ~1.9 K below 1 GPa from the superconducting transition in orthorhombic $UTe_2$ at low pressures but were not able to detect an obvious transition in the pressure region where $UTe_2$ has the tetragonal structure. However, $R(T)$ data at high pressure, as presented in Fig. 1, clearly demonstrates dramatic changes in electrical transport and the superconducting transition above ~4 GPa where $UTe_2$ transforms into the tetragonal phase. We also noted that for pressures near this transition, the cooling $R(T)$ curves deviate from the warming ones above ~80 K, which is likely because $UTe_2$ is undergoing the orthorhombic-to-tetragonal transition promoted by the increase in pressure that occurs in the DAC upon cooling. Before the structural transition, $R(T)$ in the normal state exhibits Kondo-like behavior at high $T$ with a pronounced peak at ~10 K; after the transition, the $R(T)$ curves are dramatically different than those before the transition and feature metallic behavior ($dR(T)/dT > 0$) with a knee at $T^{**} \approx 235$ K which we define as the temperature where the two extrapolated linear regions near the knee intersect. The curve at ~ 4.8 GPa near the boundary of the structural transition shows a dual nature in which the peak in $R$ (orthorhombic feature) and the knee in $R$ accompanied by the superconducting transition (tetragonal features) coexist. Interestingly, during release of pressure, $R(T)$ at 1.0 and 0.4 GPa retains the knee but there is no drop in resistance due to the superconducting transition, indicating the metastable tetragonal phase that persists to nearly ambient pressure does not exhibit superconductivity above 1.2 K.

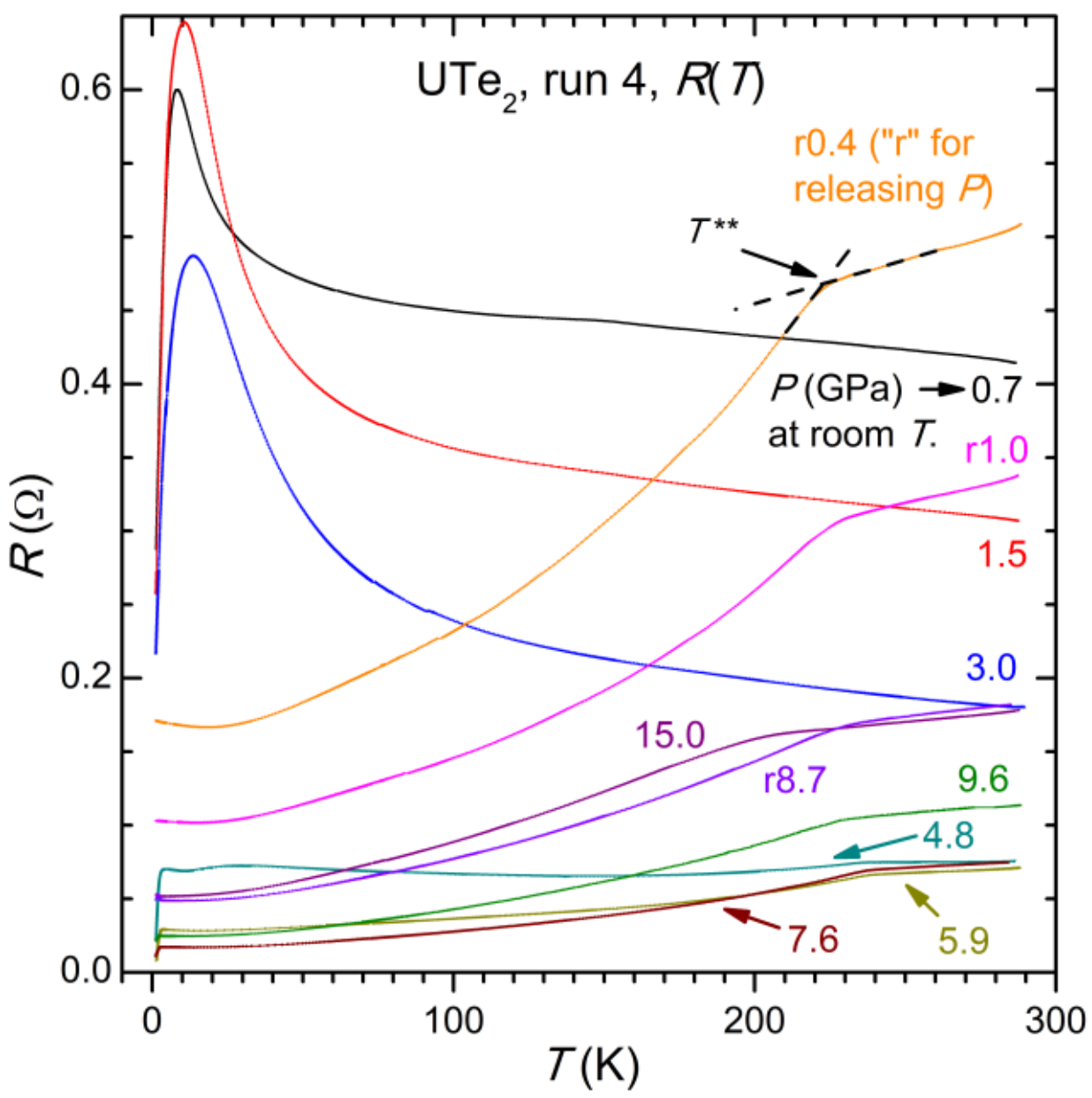

Figure1. Representative $R(T)$ curves for $UTe_2$ measured at different pressures in Run 4. The values of pressure shown were determined at room temperature. The letter “r” before some of the numbers refers to measurements taken upon release of pressure from the highest pressure of 15 GPa.

To more clearly display the superconducting transition of tetragonal $UTe_2$, Fig. 2 shows $R(T)$ curves in the range 1 K to 6 K for $P > 4$ GPa. For most of the curves, a clear resistance drop indicates the occurrence of a superconducting transition which confirms the report by Honda *et al.* [Honda23], although zero resistance could not be achieved due to the low temperature limit of the experiments and, possibly, pressure inhomogeneity (stress) across the sample manifested in the broad transition width. Above 12 GPa, an unexpected resistance increase instead of decrease with cooling was observed, which could be attributed to the redistribution of electrical current that often appears during superconducting transitions in inhomogeneous-pressurized samples. As a result, we propose that to the highest pressure we investigated, namely about 16 GPa, the tetragonal $UTe_2$ is still superconducting with a $T_c$ above 1 K.

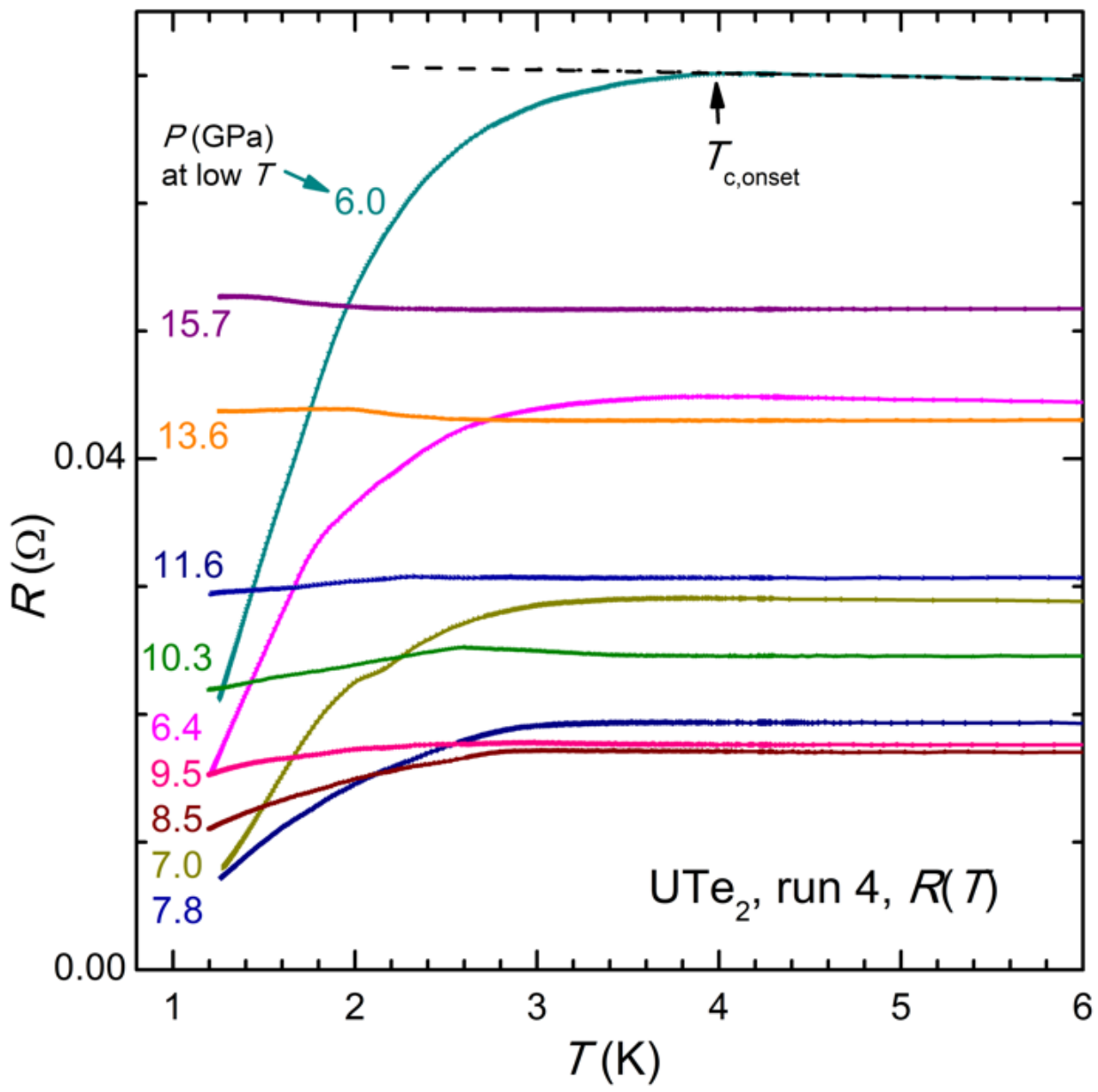

Figure 2. Resistive superconducting transition curves of the tetragonal $UTe_2$ sample studied in run 4. The $R(T)$ data are for pressures calibrated at temperatures near the superconducting transitions. $T_{c,onset}$ is defined as the temperature where the $R(T)$ curve starts to deviate from the linear extrapolation of $R(T)$ in the normal state.

Magnetic fields were applied to the tetragonal sample to study the upper critical field ($H_{c2}$) of $UTe_2$ at r5.3 GPa. As shown in Fig. 3, a field as low as 1 T almost completely suppressed the superconducting transition. Based on the Werthamer-Helfand-Hohenberg (WHH) theory, $H_{c2}(0)$ due to the orbital pair-breaking mechanism in the dirty limit can be estimated using the equation:

$$H_{c2}(0) = -0.693\, T_c(0) \frac{dH_{c2}}{dT_c}\Big|_{T_c(0)} \qquad \text{Eqn. 1}$$

We obtain $H_{c2}(0) \approx 1.2$ T for tetragonal $UTe_2$ at r5.3 GPa. In contrast, for $UTe_2$ in the orthorhombic structure, the smallest $H_{c2}(0)$ which is along the magnetic easy axis (a-axis) is about 5 T [Ran19], and even larger (10 T) [Wu24] in $UTe_2$ with enhanced superconductivity that was synthesized by a similar method as our sample. On the other hand, $H_{c2}(0)$ dominated by the paramagnetic pair-breaking mechanism (Pauli limit) can be calculated using the equation:

$$H_{Pauli} = 1.84\, T_c(0) \qquad \text{Eqn. 2}$$

Using the data at r5.3 GPa, the Pauli limit is ~ 5.1 T, significantly larger than the orbital limit, which suggests the superconductivity is mainly orbitally limited, different from the situation of $UTe_2$ at ambient pressure (in the orthorhombic phase). The Ginzburg-Landau coherence length $\xi$ can be estimated from the relation:

$$\mu_0 H_{c2}(0) = \Phi_0 / 2\pi\xi^2 \qquad \text{Eqn. 3}$$

where $\Phi_0$ is the magnetic flux quantum, which yields for tetragonal $UTe_2$ the value $\xi_{tetra} \approx 16.5$ nm. Note that the orientation of the pressurized $UTe_2$ samples was not determined so this value is an orientational average not specified for a certain axis of tetragonal $UTe_2$.

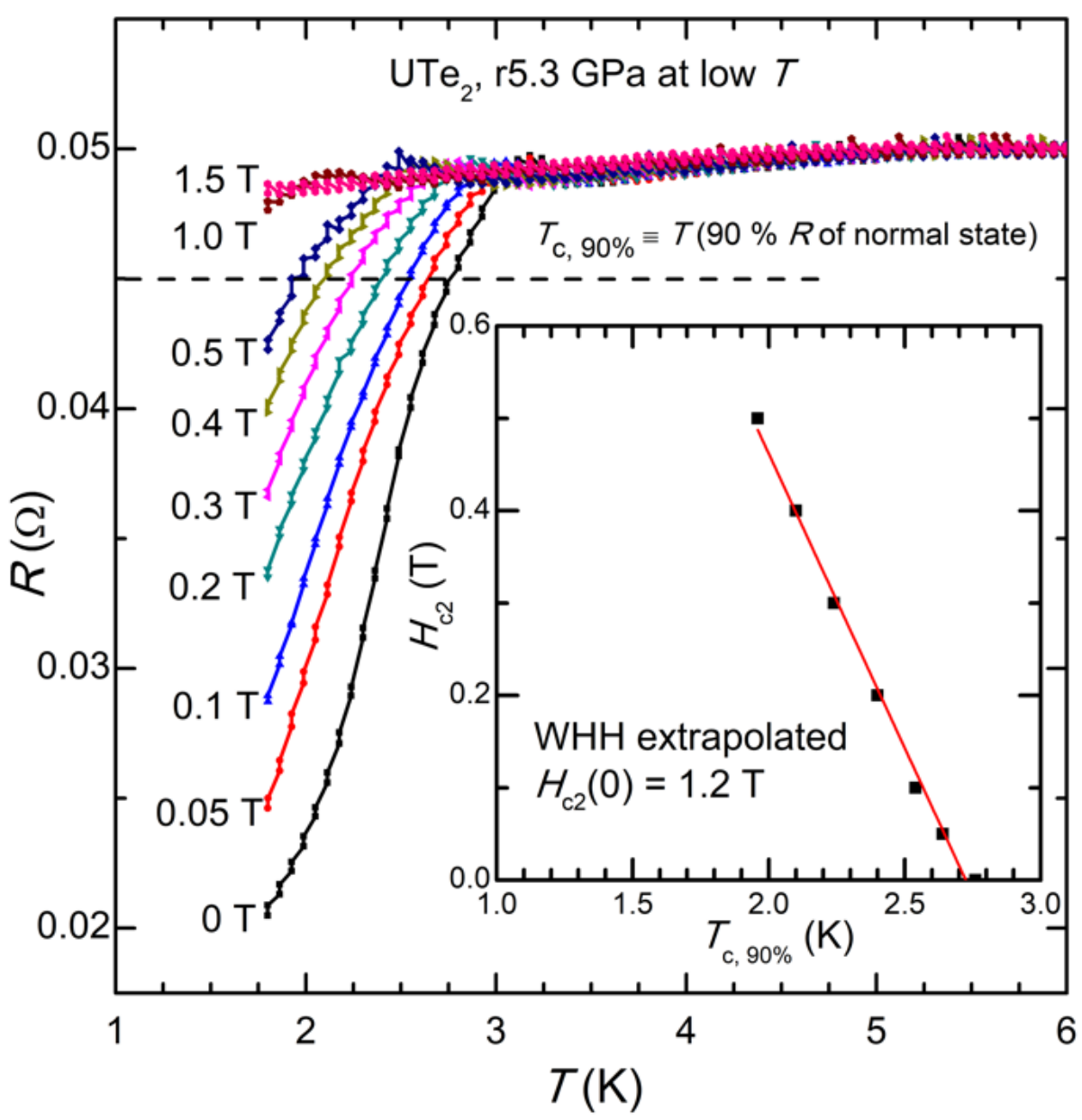

Figure 3. $R(T)$ curves measured in magnetic fields from 0 T to 1.5 T at $P$ = r5.3 GPa. $T_{c,\,90\%}$ was determined from the temperature at which $R$ drops to 90% of its value in the normal state. The inset shows a plot of the upper critical field $H_{c2}$ vs $T_{c,90\%}$ in low fields. The red straight line is a linear fit to the data which was used to estimate $H_{c2}(0)$ by means of the Werthamer-Helfand-Hohenberg (WHH) theory.

A $T$–$P$ phase diagram was constructed as shown in Fig. 4 that reveals the evolution of superconductivity and the knee in $R(T)$ at $T^{**}$ in tetragonal $UTe_2$ with applied pressure – the green points are taken from Honda *et al.* [Honda23] and the closed points (collected during increasing $P$) and half-closed points/bold face upper error bars (collected during releasing $P$) from our $R(T,P)$ and AC $\chi(T,P)$ measurements. Blue shading indicates the superconducting $T$–$P$ region which mainly resides within the tetragonal phase of $UTe_2$. The small shaded region to the left of the green line delineating the orthorhombic-tetragonal phase boundary probably belongs in the tetragonal region and is displaced due to a small differences in pressure scales between our experiments and those of Honda *et al*. [Honda23], unavoidable pressure gradients in our DAC experiments, uncertainty in determining the low-$T$ pressure, and the sluggish nature of the structural transition [Huston22, Deng24]. As can be seen, the tetragonal superconductivity ($SC_{tetra}$) could extend to ~18 GPa, taking a "half dome" shape with a broad maximum near the orthorhombic-tetragonal phase boundary.

The resistance knee is an important feature of the tetragonal phase of $UTe_2$ and appears simultaneously with the tetragonal phase superconductivity. The knee feature is very robust and persists to the highest pressure we investigated: $T^{**}(P)$ increases slightly from 237 K at 4.8 GPa to 238 K at 5.9 GPa and then decreases slowly with pressure to 204 K at 15 GPa. There may be a correlation between $T^{**}$ and $T_c$ in the tetragonal phase of $UTe_2$, namely, the maximum $T^{**}$ is at pressures where $T_c$ reaches nearly its highest value. This correlation was also observed in Ref. [Honda23] where more hydrostatic pressure was applied. Although $T^{**}$ becomes a little smaller with decreasing $P$, the resistance knee remains down to nearly ambient pressure, which again shows that the knee is an electronic hallmark of the tetragonal phase as our high-pressure XRD results [SI] indicated that the tetragonal phase is retained at least down to 1.4 GPa.

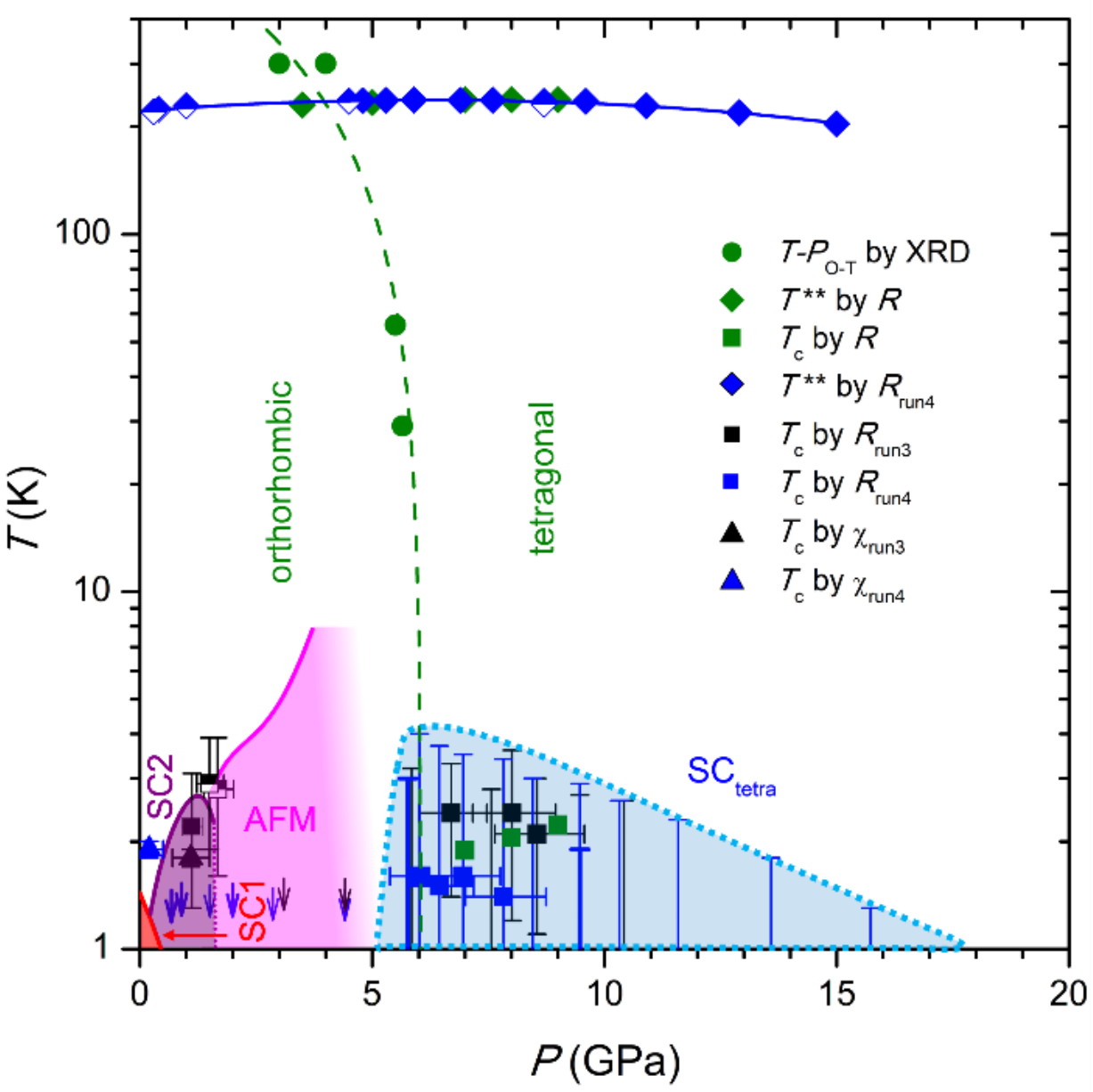

Figure 4. $T$–$P$ phase diagram for $UTe_2$. For $R(T,P)$ measurements, $T_{c,zero}$ (lower error bar) is determined by extrapolating the linear region to $R = 0\ \Omega$. $T_{c,mid}$ (symbols) is determined by finding the temperature at the midpoint between the resistances at $T_{c,onset}$ (upper end of error bar) and $T_{c,zero}$. For some curves, the extrapolated $T_{c,zero} < 0$ K, so $T_{c,zero}$ and $T_{c,mid}$ could not be determined. Down pointing arrows indicate no superconductivity was found down to $T$ at the point of the arrow. For AC $\chi(T,P)$ measurements, the triangular symbols and their error bars stand for the midpoints and the onset/end point temperatures of the transitions, respectively. The dashed line through the open green circles [Honda23] separates the orthorhombic and tetragonal structure regions. The orthorhombic $UTe_2$ superconductivity regions ($SC_{1,2}$) are depicted based on data reported in Refs. [Knafo25, Thomas20, Honda23].

*The possible origin of the transition at $T^{**}$*

As discussed above, the resistance knee at $T^{**}$ seems to be a hallmark of the electrical transport of tetragonal $UTe_2$. It resembles the knee features at $T^{*} \approx 50$ K in the $R(T)$ curves of $UTe_2$ along the $a$ and $b$ axes at ambient pressure [Ran19, Eo22] which are usually attributed to a transition from single-ion Kondo behavior to a coherent heavy fermion ground state, but it occurs at a significantly higher temperature (about 4–5 times higher). There are some other reports concerning the $T^{**}$ transition [Vališka21, Honda23, Thebault24] in which a Bridgman anvil cell or a cubic anvil cell was used, providing a more hydrostatic pressure environment than our diamond anvil cell, suggesting the transition is not solely due to the effect of strain. On the other hand, it was found that this transition appeared when the $b$-axis of an orthorhombic $UTe_2$ crystal was parallel to the direction of applied pressure but was absent for a sample with the $c$-axis aligned along the direction of applied pressure [Vališka21], revealing the role of uniaxial pressure in inducing the transition. If the resistance knee signals the transition to a coherent ground state, it indicates that the Kondo interaction is stronger in tetragonal $UTe_2$ than in orthorhombic $UTe_2$ and thus enters the coherent ground state at a much higher $T^{**}$ than in orthorhombic $UTe_2$. The stronger Kondo interaction indicates larger hybridization between the U-5$f$- and conduction-electron states and a smaller conduction electron effective mass $m^{*}$, consistent with a smaller coefficient of the Fermi liquid contribution to the electrical resistivity $A = (\rho - \rho_o)/T^2$ ($\rho_o$ is the residual resistivity) at low temperatures $T \ll T^{**}$, which is about one thousandth of that of the orthorhombic phase [Honda23].

Before the establishment of a coherent ground state, the electrical resistivity $\rho(T)$ should have a contribution due to incoherent Kondo scattering that varies as ~ $-\ln T$ and increases in magnitude upon cooling. This contribution was not evident in our measurements (orientation of measured $UTe_2$ not determined) or those reported in Refs. [Honda23, Thebault24] (resistance measured along the *a*-axis of the orthorhombic sample). However, it should be noted that the $-\ln T$ contribution could be difficult to observe because it is based on a single ion theory and could be modified by interactions between U ions, uncertainty in determining the Kondo temperature, and difficulties in separating it from the lattice contribution to $\rho(T)$. An alternate interpretation of the apparent transition at $T^{**}$ was proposed by Thebault *et al*. [Thebault24]. These authors measured the electrical resistivity of $UTe_2$ in the tetragonal phase in strong magnetic fields and observed step functions in the $\rho(H)$ curves at $H_{x1}$ and $H_{x2}$, only at temperatures below $T^{**}$, which lead the authors to propose that $T^{**}$ could be the temperature of a magnetic phase transition below which the spin disorder contribution to electron scattering decreases, and $H_{x1,2}$ could be the signatures of some metamagnetic transitions in magnetically ordered tetragonal $UTe_2$ [Thebault24]. According to high-pressure XRD results [Huston22, Deng24], below 30 GPa, the nearest U-U distance in the tetragonal phase, $d_{U\text{-}U}$, is larger than the Hill limit (~3.5 Å) [Hill70] for U compounds, favoring a more localized U-5*f* electron configuration and thus a magnetically ordered state [Mydosh11]. From this point of view, it is natural to expect tetragonal $UTe_2$ to undergo a transition to some kind of magnetically ordered state.

Although with current data we are unable to help clarify the debate on the origin of the transition at $T^{**}$, this transition is certainly associated with $SC_{tetra}$: in a certain pressure range, the state below $T^{**}$, either Kondo coherence, magnetic order or something else, and $SC_{tetra}$ coexist down to low temperatures. On the other hand, upon releasing pressure to nearly ambient pressure, even though the measured $T^{**}$ remains similar to that observed at high pressures, we did not detect $SC_{tetra}$ above 1.3 K. This decoupling indicates the transition at $T^{**}$ is determined by an electronic origin that depends weakly on changes in lattice parameters of the tetragonal phase, in contrast to $SC_{tetra}$ that is more sensitive to changes in the lattice parameters.

*Conventional or unconventional superconductivity in tetragonal $UTe_2$*

We obtained $\xi_{tetra} \approx 16.5$ nm for the Ginzburg-Landau coherence lengths using Eqn. 3. Orthorhombic $UTe_2$ has high magnetic anisotropy and the coherence lengths at ambient pressure were reported to vary from 2.4 nm to 15 nm [Li25, Sharma25, Yang25] depending on the direction of the applied magnetic field and measurement methods. The larger $\xi_{tetra}$ and the relation $H_{c2}(0) < H_{Pauli}$ indicate $SC_{tetra}$ is more conventional than its orthorhombic counterpart ($SC_{ortho}$). Moreover, the increased symmetry and enlarged U nearest neighbor distance $d_{U\text{-}U}$ in the tetragonal phase [Deng24] imply reduced 5*f* electron participation in bonding, suggesting a crossover toward a more itinerant electronic state, potentially favorable to more conventional BCS-like superconductivity. It is noteworthy that the element tellurium (Te) at 4.5 GPa has a $T_c$ of 4 K, $H_{c2}(0)$ of ~0.3 T, and coherence length of ~34 nm [Zhao23]. The big difference in $H_{c2}(0)$ between Te and $UTe_2$ at similar pressures indicates that the observed tetragonal superconductivity was not from decomposition of $UTe_2$ into U and Te during compression/decompression, which was found in Ref. [Huston22].

The orthorhombic phase of $UTe_2$ contains a network of two-leg ladders extending in the *a*-direction in which pairs of U atoms, or U dimers with the shortest U-U distance, form the rungs of ladders which are oriented in the *c*-direction. Based on elastic neutron scattering measurements, there is evidence for FM exchange interactions between the U ions in the dimers that form the rungs of the ladder and AFM exchange interactions between U atoms in neighboring ladders [Knafo21]. The FM intraunit cell interactions between U ions within a U dimer have been considered as possible mechanisms for spin triplet pairing in $UTe_2$ [Chen21, Shishidou21]. Chen *et al*. [Chen21] have proposed that AFM interunit cell exchange interactions could account for the incommensurate AFM fluctuations observed in inelastic neutron scattering experiments [Duan20, Duan21, Knafo21, Butch22] and the resonance in the spin spectrum of $UTe_2$ [Duan21, Raymond21], as well as the incommensurate AFM ordering under pressure [Knafo25]. In contrast, there are no U-dimers in the tetragonal phase of $UTe_2$, which could account for the fact that it exhibits a more conventional type of superconductivity, as noted above.

The discovery of $SC_{tetra}$ places U-based superconductivity in a broader and more intriguing context. Canonical heavy fermion uranium superconductors, such as orthorhombic $UTe_2$, $UPt_3$, $UBe_{13}$, and $URu_2Si_2$ [Pfleiderer09], typically exceed the Hill limit [Mydosh11], and exhibit strongly correlated electron behavior and unconventional superconducting electron pairing. In contrast, some other uranium superconductors with a smaller Sommerfeld coefficient $\gamma$, including α-U under pressure [Ho66, Maple72] and β-U [Matthias66], usually fall below the Hill limit [Mydosh11], displaying relatively more BCS-like superconductivity. Strikingly, tetragonal $UTe_2$ violates this empirical dichotomy; it has the highest $T_c \approx 4$ K among all known U-based intermetallic superconductors, its nearest $d_{U\text{-}U}$ exceeds the Hill limit, yet this $SC_{tetra}$ seems to belong to a conventional category based on the previous discussion of its structural symmetry, $H_{c2}(0)$ and *A* coefficient of $R(T)$ in the normal state. Tetragonal $UTe_2$ constitutes a rare scenario where a U-compound potentially harbors BCS-like superconductivity instead of unconventional superconductivity or 5*f* electron magnetic ordering with a large spacing between U ions, which places $UTe_2$ in a previously unexplored regime where structural, electronic, and pairing tendencies are fundamentally at odds with established trends.

We have established the *P*–*T* phase diagram for $UTe_2$ through measurements that extend to substantially higher pressures than those reported in Ref. [Honda23], revealing the complete superconducting region within the tetragonal phase of $UTe_2$ which extends up to at least 30 GPa [Deng24]. The resulting phase diagram exposes a striking dichotomy: a narrow, fragile, unconventional superconducting region associated with the orthorhombic phase, in sharp contrast to a wide, robust, potentially conventional superconducting dome stabilized in the tetragonal phase. Below ~5 GPa, multiple kinds of order, including $SC_1$, $SC_2$, and AFM, coexist or compete [Thomas20, Knafo25], whereas above this threshold, $SC_{tetra}$ rapidly dominates and remains relatively stable over an extended pressure range. These observations raise a central and unresolved question: can pressure continuously tune between BCS-like and unconventional superconducting regimes within a single material, like $UTe_2$? If so, $UTe_2$ would constitute a rare platform for directly interrogating the microscopic evolution between distinct pairing mechanisms though the application of pressure. Our results therefore position $UTe_2$ as a critical bridge connecting conventional and unconventional superconductivity, offering a new route to unify these seemingly disparate regimes within a single correlated electron system.

## Concluding Remarks

In summary, through electrical transport and AC magnetic susceptibility measurements, we have constructed the *P–T* phase diagram of $UTe_2$ and identified distinct superconducting regions associated with the orthorhombic and tetragonal phases. Superconductivity in the tetragonal phase is relatively robust with respect to the application of pressure and exhibits a lower upper critical field and larger coherence length, consistent with more conventional superconductivity with orbitally limited behavior. These results suggest a crossover toward a more weakly correlated superconducting state involving fewer 5*f*-electrons under pressure, highlighting the strong connection between structure, electronic correlations, and superconductivity in the $UTe_2$ system, that can serve as a testing ground for pressure tuned superconducting mechanisms.

**Data, Materials, and Software Availability.** All study data are included in the article and/or SI Appendix [SI], are available in the OSF digital repository [Data].

**Acknowledgement**

Research at the University of California, San Diego was supported by the National Nuclear Security Administration (NNSA) under the Stewardship Science Academic Alliance Program through the US DOE under Grant DE-NA0004235 (electrical transport measurements at high pressure), and the US Department of Energy (DOE) Basic Energy Sciences (BES) under Grant DE-FG02-04ER46105 ($UTe_2$ single crystal growth and characterizations at ambient pressure). Research at Michigan State University (high-pressure X-ray diffraction) was supported by the U.S. DOE BES under Contract DE-SC0023648.